\documentclass[useAMS,usenatbib]{mn2e}

\usepackage{xspace}
\usepackage{graphicx}
\newcommand{\ignore}[1]{}
\providecommand{\ao}{}
\renewcommand{\ao}{adaptive optics (AO)\renewcommand{\ao}{AO\xspace}\renewcommand{\Ao}{AO\xspace}\xspace}
\newcommand{\Ao}{Adaptive optics (AO)\renewcommand{\ao}{AO\xspace}\renewcommand{\Ao}{AO\xspace}\xspace}
\newcommand{\wfs}{wavefront sensor (WFS)\renewcommand{\wfs}{WFS\xspace}\renewcommand{\wfss}{WFSs\xspace}\xspace}
\newcommand{\wfss}{wavefront sensors (WFSs)\renewcommand{\wfs}{WFS\xspace}\renewcommand{\wfss}{WFSs\xspace}\xspace}
\newcommand{\shwfs}{Shack-Hartmann \wfs (SHWFS)\renewcommand{\shwfs}{SHWFS\xspace}\xspace}
\newcommand{\dm}{deformable mirror (DM)\renewcommand{\dm}{DM\xspace}\renewcommand{\dms}{DMs\xspace}\renewcommand{\Dms}{DMs\xspace}\renewcommand{\Dm}{DM\xspace}\xspace}
\newcommand{\dms}{deformable mirrors (DMs)\renewcommand{\dm}{DM\xspace}\renewcommand{\dms}{DMs\xspace}\renewcommand{\Dms}{DMs\xspace}\renewcommand{\Dm}{DM\xspace}\xspace}
\newcommand{\Dms}{Deformable mirrors (DMs)\renewcommand{\dm}{DM\xspace}\renewcommand{\dms}{DMs\xspace}\renewcommand{\Dms}{DMs\xspace}\renewcommand{\Dm}{DM\xspace}\xspace}
\newcommand{\Dm}{Deformable mirror (DM)\renewcommand{\dm}{DM\xspace}\renewcommand{\dms}{DMs\xspace}\renewcommand{\Dms}{DMs\xspace}\renewcommand{\Dm}{DM\xspace}\xspace}

\newcommand{\shs}{Shack-Hartmann sensor (SHS)\renewcommand{\shs}{SHS\xspace}\renewcommand{\shss}{SHSs\xspace}\xspace}
\newcommand{\shss}{Shack-Hartmann sensors (SHSs)\renewcommand{\shs}{SHS\xspace}\renewcommand{\shss}{SHSs\xspace}\xspace}
\newcommand{\lgs}{laser guide star (LGS)\renewcommand{\lgs}{LGS\xspace}\renewcommand{\lgss}{LGSs\xspace}\xspace}
\newcommand{\lgss}{laser guide stars (LGSs)\renewcommand{\lgs}{LGS\xspace}\renewcommand{\lgss}{LGSs\xspace}\xspace}
\newcommand{\ngs}{natural guide star (NGS)\renewcommand{\ngs}{NGS\xspace}\renewcommand{\ngss}{NGSs\xspace}\xspace}
\newcommand{\ngss}{natural guide stars (NGSs)\renewcommand{\ngs}{NGS\xspace}\renewcommand{\ngss}{NGSs\xspace}\xspace}
\newcommand{\mems}{Micro-Electro-Mechanical Systems (MEMS)\renewcommand{\mems}{MEMS\xspace}\xspace}
\newcommand{\snr}{signal to noise ratio (SNR)\renewcommand{\snr}{SNR\xspace}\xspace}
\newcommand{\moao}{multi-object \ao (MOAO)\renewcommand{\moao}{MOAO\xspace}\xspace}
\newcommand{\mcao}{multi-conjugate adaptive optics (MCAO)\renewcommand{\mcao}{MCAO\xspace}\xspace}
\newcommand{\ltao}{laser tomographic adaptive optics (LTAO)\renewcommand{\ltao}{LTAO\xspace}\xspace}
\newcommand{\cpu}{central processing unit (CPU)\renewcommand{\cpu}{CPU\xspace}\renewcommand{\cpus}{CPUs\xspace}\xspace}
\newcommand{\cpus}{central processing units (CPUs)\renewcommand{\cpu}{CPU\xspace}\renewcommand{\cpus}{CPUs\xspace}\xspace}
\newcommand{\psf}{point spread function (PSF)\renewcommand{\psf}{PSF\xspace}\renewcommand{\psfs}{PSFs\xspace}\xspace}
\newcommand{\psfs}{point spread functions (PSFs)\renewcommand{\psf}{PSF\xspace}\renewcommand{\psfs}{PSFs\xspace}\xspace}
\newcommand{\fpga}{field programmable gate array (FPGA)\renewcommand{\fpga}{FPGA\xspace}\renewcommand{\fpgas}{FPGAs\xspace}\xspace}
\newcommand{\fpgas}{field programmable gate arrays (FPGAs)\renewcommand{\fpga}{FPGA\xspace}\renewcommand{\fpgas}{FPGAs\xspace}\xspace}
\newcommand{\sor}{successive over-relaxation (SOR)\renewcommand{\sor}{SOR\xspace}\xspace}
\newcommand{\fdpcg}{Fourier domain pre-conditioned gradient (FDPCG)\renewcommand{\fdpcg}{FDPCG\xspace}\xspace}
\newcommand{\map}{maximum a-posteriori (MAP)\renewcommand{\map}{MAP\xspace}\xspace}
\newcommand{\elt}{Extremely Large Telescope (ELT)\renewcommand{\elt}{ELT\xspace}\renewcommand{\elts}{ELTs\xspace}\xspace}
\newcommand{\elts}{Extremely Large Telescopes (ELTs)\renewcommand{\elt}{ELT\xspace}\renewcommand{\elts}{ELTs\xspace}\xspace}
\newcommand{\dugall}{Durham University generalised adaptive optics laser laboratory (DUGALL)\renewcommand{\dugall}{DUGALL\xspace}\xspace}
\newcommand{\fwhm}{full-width at half-maximum (FWHM)\renewcommand{\fwhm}{FWHM\xspace}\xspace}
\newcommand{\wht}{William Herschel Telescope (WHT)\renewcommand{\wht}{WHT\xspace}\xspace}
\newcommand{\emccd}{electron multiplying CCD (EMCCD)\renewcommand{\emccd}{EMCCD\xspace}\xspace}
\newcommand{\dasp}{Durham \ao simulation platform (DASP)\renewcommand{\dasp}{DASP\xspace}\xspace}
\newcommand{\eelt}{European \elt (E-ELT)\renewcommand{\eelt}{E-ELT\xspace}\xspace}
\newcommand{\mpi}{Message Passing Interface (MPI)\renewcommand{\mpi}{MPI\xspace}\xspace}
\newcommand{\smp}{symmetric multi-processing (SMP)\renewcommand{\smp}{SMP\xspace}\xspace}
\newcommand{\svd}{singular value decomposition (SVD)\renewcommand{\svd}{SVD\xspace}\xspace}
\newcommand{\gpu}{graphical processing unit (GPU)\renewcommand{\gpu}{GPU\xspace}\renewcommand{\gpus}{GPUs\xspace}\xspace}
\newcommand{\gpus}{graphical processing units (GPUs)\renewcommand{\gpu}{GPU\xspace}\renewcommand{\gpus}{GPUs\xspace}\xspace}
\newcommand{\fft}{fast Fourier transform (FFT)\renewcommand{\fft}{FFT\xspace}\xspace}
\newcommand{\ifu}{integral field unit (IFU)\renewcommand{\ifu}{IFU\xspace}\xspace}
\newcommand{\darc}{the Durham \ao real-time controller (DARC)\renewcommand{\darc}{DARC\xspace}\renewcommand{\Darc}{DARC\xspace}\xspace}
\newcommand{\Darc}{The Durham \ao real-time controller (DARC)\renewcommand{\darc}{DARC\xspace}\renewcommand{\Darc}{DARC\xspace}\xspace}
\newcommand{\cots}{commercial off-the-shelf (COTS)\renewcommand{\cots}{COTS\xspace}\xspace}
\newcommand{\rtcp}{real-time control pipeline (RTCP)\renewcommand{\rtcp}{RTCP\xspace}\xspace}
\newcommand{\rms}{root-mean-square (RMS)\renewcommand{\rms}{RMS\xspace}\xspace}
\newcommand{\sFPDP}{serial Front Panel Data Port (sFPDP)\renewcommand{\sFPDP}{sFPDP\xspace}\xspace}
\newcommand{\wpu}{wavefront processing unit (WPU)\renewcommand{\wpu}{WPU\xspace}\xspace}
\newcommand{\canary}{CANARY\xspace}
\newcommand{\Canary}{CANARY\xspace}
\newcommand{\rtcs}{real-time control system (RTCS)\renewcommand{\rtcs}{RTCS\xspace}\xspace}

\title[Wavefront sensing with brightest pixel selection]{Wavefront
  sensing with a brightest pixel selection algorithm}

\author[A. G. Basden, R. M. Myers and
  E. Gendron]{A. G. Basden$^{1}$\thanks{E-mail: a.g.basden@durham.ac.uk (AGB)}, R. M. Myers$^1$ and E. Gendron$^{2}$\\
$^{1}$Department of Physics, South Road, Durham, DH1 3LE, UK\\
$^{2}$Observatoire de Paris, Meudon, Paris, France}
\begin{document}
\maketitle


\begin{abstract}
Astronomical adaptive optics systems with open-loop deformable mirror
control have recently come on-line.  In these systems, the deformable
mirror surface is not included in the wavefront sensor paths, and so
changes made to the deformable mirror are not fed back to the
wavefront sensors.  This gives rise to all sorts of linearity and
control issues mainly centred on one question: Has the mirror taken
the shape requested?  Non-linearities in wavefront measurement and in
the deformable mirror shape can lead to significant deviations in
mirror shape from the requested shape.  Here, wavefront sensor
measurements made using a brightest pixel selection method are
discussed along with the implications that this has for open-loop AO
systems.  Discussion includes elongated laser guide star spots and
also computational efficiency.

\end{abstract}

\section{Introduction}
\Ao is a widely used technology on large near infra-red and optical
telescopes, and almost all current and planned facility class science
telescopes will have an \ao system.  \Ao systems have been used to obtain a
large number of science results that would otherwise have been
impossible due to atmospheric induced
perturbations\citep{2004A&A...417L..21G,2005ApJ...625.1004M}.  Novel
applications for \ao such as high resolution imaging over a wide-field
are currently being studied and implemented\citep{2004SPIE.5490..236Mshort},
for example \moao \citep{canaryresultsshort}.

Starlight passing through the Earth's atmosphere is distorted by the
introduction of random perturbations due to time-varying refractive
index changes in the atmosphere\citep{tatarski}.  This means that
forming a diffraction limited image is no longer possible, and the
effective resolution of a telescope is reduced.  However, corrective
measures can be taken, which by changing the shape of a \dm in response
to a \wfs can greatly reduce the impact of the atmospheric turbulence
\citep{roddier}.  An \ao system, used to perform this task, is composed
of \wfss and \dms.

Traditional \ao systems use a \wfs to view the incident wavefront
after correction by the \dm has been applied, meaning that only
partial further correction is then required in the resulting
closed-loop system.  However, recent progress has been made with
open-loop systems \citep{canaryresultsshort} such as those required by \moao.
In these systems, the wavefront sensors are not sensitive to the
corrections applied to the \dm, but rather measure the perturbed
wavefront directly.  These measurements are then used to control the
\dm without subsequent direct measurement of whether the desired control
shape has been realised (so called open-loop control), reducing the
impact of atmospheric turbulence for the science path only, but not
for the \wfss.  In such open-loop systems, the characteristics of the
\dm and \wfss must be well known, as any non-linearities will result
in the wrong correction being applied to the \dm.  It is therefore
desirable that any processing algorithm used should be linear.

A Shack-Hartmann wavefront sensor consists of an array of lenses, each
of which effectively samples part of the telescope pupil plane.  The
local tilt of a wavefront across each lenslet is measured by
determining the position of the spot produced by the lenslet on the
detector; if there is no tilt, the spot on the detector will appear in
a position corresponding to the centre of the sub-aperture.  The
standard way of determining the spot position, which is used in almost
all Shack-Hartmann based astronomical \ao systems used on-sky, is by using a
centre of gravity algorithm \citep{2006MNRAS.371..323T}.  However this
is sensitive to noise (background, readout and photon arrival
statistics) and can perform poorly for open-loop systems.  

Correct image calibration (including removal of background light and
detector dark noise, application of a flat field map and thresholding)
is very difficult to achieve, as the necessary calibration maps may be
time and temperature dependent.  Incorrect calibration leads to
incorrect wavefront estimation and hence to incorrect correction with
a \dm, which for open-loop systems can lead to significant performance
loss.  This is less of a problem for closed-loop systems because spot
motions are usually small and less affected by these systematic errors
as static aberrations can be corrected using \wfs offsets (reference
measurements).  However incorrect calibration may lead to incorrect
\wfs offsets, giving rise to non-common path wavefront error.
Detector noise, and noise due to the random stochastic nature of
photon arrival times will also impact wavefront estimation accuracy.
For typical \ao system working regimes using faint fluxes, this is a
significant source of noise.

Further, when a \lgs is used, the shape of Shack-Hartmann spots
changes across the wavefront sensor, taking an elongated shape in
sub-apertures away from the \lgs launch position due to system
geometry.  A conventional flat threshold is then no longer ideal
\citep{2009MNRAS.398.1461L} and using a non-constant threshold can be
of advantage.

In this paper, a brightest pixel selection algorithm is investigated
which can improve wavefront slope estimation accuracy by applying a
different threshold value on a per-sub-aperture basis, and hence
improve \ao system performance.  This algorithm has previously been
proposed  (by author E.~Gendron) but never formally published or
studied, and here we report on the first real-time implementation of
it, and investigate linearity and how to best apply this algorithm in
different conditions and situations.  This algorithm is primarily
suited to Shack-Hartmann systems with at least $4\times4$ pixels per
sub-aperture.

If \S2 the brightest pixel selection algorithm is described, in \S3
performance measurements and comparisons are given, and in \S4
conclusions are made.

\section{The brightest pixel selection algorithm}
The brightest pixel algorithm as defined in this paper is simple:
Select the brightest $N$ pixels within a given Shack-Hartmann
sub-aperture, and set all other pixel values to zero.  This modified
sub-aperture is then processed in the standard way to compute
wavefront slope, for example using a centre of gravity algorithm to
measure the spot position thus obtaining the wavefront slope across
the sub-aperture.  This algorithm benefits from the removal of pixels
that contain noise only and not useful signal, and is less sensitive
to random noise than the radial thresholding technique proposed by
\citep{2009MNRAS.398.1461L} which sets a threshold at a fraction of the
brightest pixel in a sub-aperture.  To be valuable for open-loop \ao
systems, the linearity and performance of the brightest pixel
selection algorithm must be investigated.

\begin{figure}
\includegraphics[width=2.5cm]{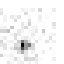}
\includegraphics[width=2.5cm]{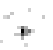}
\includegraphics[width=2.5cm]{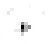}
\caption{A figure demonstrating brightest pixel selection algorithm
  for a single Shack-Hartmann sub-aperture spot with (from left to
  right) no selection, after selection of the brightest 20 pixels
  (setting all others to zero), and after selection of brightest 20
  pixels and subtraction of 21$^{\textrm{st}}$ brightest pixel,
  setting negative values to zero.  A linear scale has been used and
  each image has been normalised with pixel values ranging from 0 to 1.}
\label{fig:shs}
\end{figure}

A modified version of the brightest pixel selection algorithm is also
possible, and includes subtraction of the value of the $N+1^{th}$
brightest pixel in a given sub-aperture from all pixels in the
sub-aperture, after which all negative values are set to zero.  In this
paper, the performance of both of these algorithms is investigated.
Fig.~\ref{fig:shs} demonstrates these algorithms.

\subsection{Application of brightest pixel selection}
The brightest pixel selection algorithm has been used on-sky with the
\canary \citep{canaryshort} \ao test-bench system on the \wht in September and
November 2010.  For this purpose, the algorithm had been integrated
with the \darc \rtcs \citep{basden9}.  However the operation of this
algorithm had not been
studied in depth and so the number of pixels selected was somewhat
arbitrary, usually around 15.  In this paper, an in-depth study of the
performance of this algorithm is made, allowing more educated
determination of the number of pixels to select when used in future, for
example during the future phases of \canary in 2011--2015.

The implementation of this algorithm in the real-time control system
is fairly straightforward, though must be efficient so that system
latencies are not significantly increased, which would lead to
performance degradation.  It is carried out as follows on a
sub-aperture by sub-aperture basis.

\begin{enumerate}
\item A quick-sort algorithm is used to sort all the
  pixels of a given sub-aperture.
\item The $n^{th}$ brightest pixel is selected as the threshold
  value (the first pixel being the brightest).
\item The $n+1^{th}$ brightest pixel is selected as an optional
  offset (bias) value.
\item The sub-aperture pixels with intensity below the threshold value
  are set to zero.
\item Optionally, pixels with intensity equal to or greater than the
  threshold value have the offset (bias) value subtracted.  We
  hereafter term the version of this algorithm which includes this
  subtraction the ``$N+1^{th}$ subtracted algorithm''.
\item Shack-Hartmann slope measurement then proceeds in a standard way
  (e.g.\ weighted centre of gravity, correlation, etc).
\end{enumerate}

Sorting the pixels in each sub-aperture is computationally expensive,
particularly for systems with larger numbers of pixels per
sub-aperture.  \Canary has 256 pixels per \ngs sub-aperture, though
fortunately, the \rtcs used is powerful enough to handle these extra
calculations without adding significant latency.  In
section~\ref{sect:computational} the computational complexity of the
brightest pixel selection algorithm is explored.

\subsection{Performance criteria}
\label{sect:thealgo}
The performance of the pixel selection algorithm has been investigated
using Monte-Carlo simulation of an open-loop Shack-Hartmann wavefront
sensor (i.e., viewing the atmospheric turbulence only, not a
deformable mirror) following \citep{basden5}.  The slope measurements
computed when using the brightest pixel selection algorithm are
compared with slope measurements computed using a noiseless image
without the brightest pixel selection algorithm (giving the best
Shack-Hartmann estimation of the true slope measurement), and the \rms
error of many thousands of such comparisons is computed.  For each
comparison a different realisation of atmospheric turbulence is used.
In this paper, the performance criteria is therefore defined by

\begin{equation}
R = \sqrt{\frac{\sum_{m=1}^N
    \left(S_\textrm{true}(m)-S_\textrm{estimated}(m)  \right)^2}{N}}
\label{eq:metric}
\end{equation}
where $N$ is the number of measurements taken and $S(m)$ is the
$m^{th}$ individual slope measurement measured with $m^{th}$
atmospheric turbulence realisation computed with either no pixel
selection and no noise (true), or with brightest pixel selection
(estimated).  A lower value of $R$ means more accurate estimation of
the local wavefront slope, i.e.\ better performance.  It should be
noted that this \rms performance criteria does not contain bias, but
is due solely to the noise introduced (photon shot noise and readout
noise).  The following operations are performed:

\begin{enumerate}
\item A realisation of the optical wavefront phase introduced by
  atmospheric perturbations is produced.
\item A \wfs is modelled producing a Shack-Hartmann spot pattern with
  photon noise and readout noise added.
\item Image calibration including brightest pixel selection is
  performed.
\item A centre of gravity algorithm is used to estimate wavefront slope
  for each sub-aperture.
\item The estimated wavefront slope is compared with the slope
  calculated using a noiseless Shack-Hartmann image (no read noise or
  photon shot noise).
\item The process is repeated many times computing the \rms difference between
  estimated and true wavefront slopes.
\end{enumerate}

\begin{figure}
\includegraphics[width=7cm]{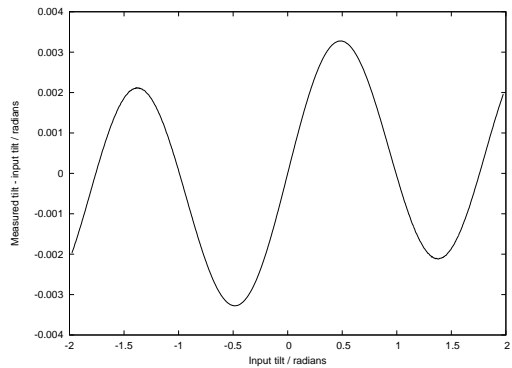}
\caption{A figure showing the bias in wavefront slope measurement by a
  noiseless Shack-Hartmann wavefront sensor.  The wavefront sensor
  here used $16\times16$ pixels per sub-aperture.  This bias is due to
  the pixellated and truncated nature of a Shack-Hartmann sensor.
}
\label{fig:perfectwfs}
\end{figure}

It should be noted that we are comparing measurements with a noiseless
Shack-Hartmann image rather than the true slope measurement.  This is
intentional, because it allows us to investigate the best possible way
of processing Shack-Hartmann data using the brightest pixel selection
algorithm, rather than comparing how well this performs relative to a
perfect wavefront sensor (which a noiseless Shack-Hartmann is not).
For reference purposes, Fig.~\ref{fig:perfectwfs} shows the deviation
(bias) of perfect Shack-Hartmann measurements from true wavefront
slope.  

In our simulations, detectors are modelled with a uniform response,
and so a flat field image of unity is used.  We do not add any
detector dark noise (it can be minimised at the typical frame rates
that astronomical wavefront sensors are operated at) and so no bias
subtraction is necessary.  A detector background of 50 counts is added
(unless otherwise stated), and a background subtraction of 50 counts
is removed.  It should be noted that we do not subtract at a level of
$3\sigma$ above detector noise because this adds a bias in slope
estimation.  

\subsection{Model description}
In this paper, results are concentrated on a model based on the
\canary \moao demonstrator instrument because that is where the
brightest pixel selection algorithm has been used on-sky.  A
telescope with a 4.2~m diameter primary mirror and an \ao system with
\wfss having $7\times7$ sub-apertures is used with a detection
wavelength of 640~nm.

The baseline configuration for these comparisons comprises a signal
level of 200 photons per sub-aperture and a \rms readout noise of 2
photo-electrons.  Each sub-aperture consists of $16\times16$ pixels
with a pixel scale of 0.22~arcsec per pixel.  The atmosphere is
modelled with Von Karman statistics with an outer scale of 30~m and a
Fried's parameter ($r_0$) of 12.5~cm (at 500~nm, representing fairly average
atmospheric statistics) corresponding to $d/r_0 = 4.8$ where $d$ is
the effective sub-aperture diameter (the telescope pupil diameter
divided by the number of sub-apertures across it).  The
Shack-Hartmann spots are modelled as having an Airy disc \psf with a
diameter (to first minimum) of 0.7~arcsec (20\% of a sub-aperture).  A
rough estimate for signal-to-noise ratio can be made by considering
photon and readout noise, giving a value of about 6 following
\begin{equation}
\mathrm{SNR} = \frac{n_\mathrm{ph}}{\sqrt{n_\mathrm{ph}+\sigma_e^2 N_\mathrm{pix}}}
\label{eq:snr}
\end{equation}
where $n_\mathrm{ph}$ is the number of photons per sub-aperture per frame,
$\sigma_e$ is the detector read noise, and $N_\mathrm{pix}$ is the number
of pixels per sub-aperture.

\subsection{Linearity}
As mentioned previously, for open-loop \ao systems, linearity is
important for achieving good correction.  The linearity of the
brightest pixel selection algorithm has been studied by placing known,
gradually increasing, tilts across a sub-aperture, and comparing the
measured slope using all pixels and with brightest pixel selection.
In both cases, the \wfs is noiseless, with no readout noise or photon
arrival statistics added.  In an ideal case, with a linear \wfs, the
measured slope will be directly proportional to the true slope.

\section{Results}
\label{sect:results}
A parameter space including the number of brightest pixels selected,
\wfs readout noise, signal level, background level errors and
atmospheric model details has been explored to find the optimum number
of brightest pixels to use for selection, and results are presented here.

For each of these results, extrapolation to using the 256 brightest
pixels (all pixels) will give the performance that would be achieved using all
available pixels (the traditional approach), though this is not
explicitly shown to simplify the data and make the results clearer.

\subsection{Sensitivity to Shack-Hartmann PSF size}
\begin{figure}
(a)\includegraphics[width=3cm]{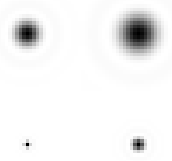}
(b)\includegraphics[width=3cm]{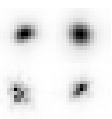}
(c)\includegraphics[width=7cm]{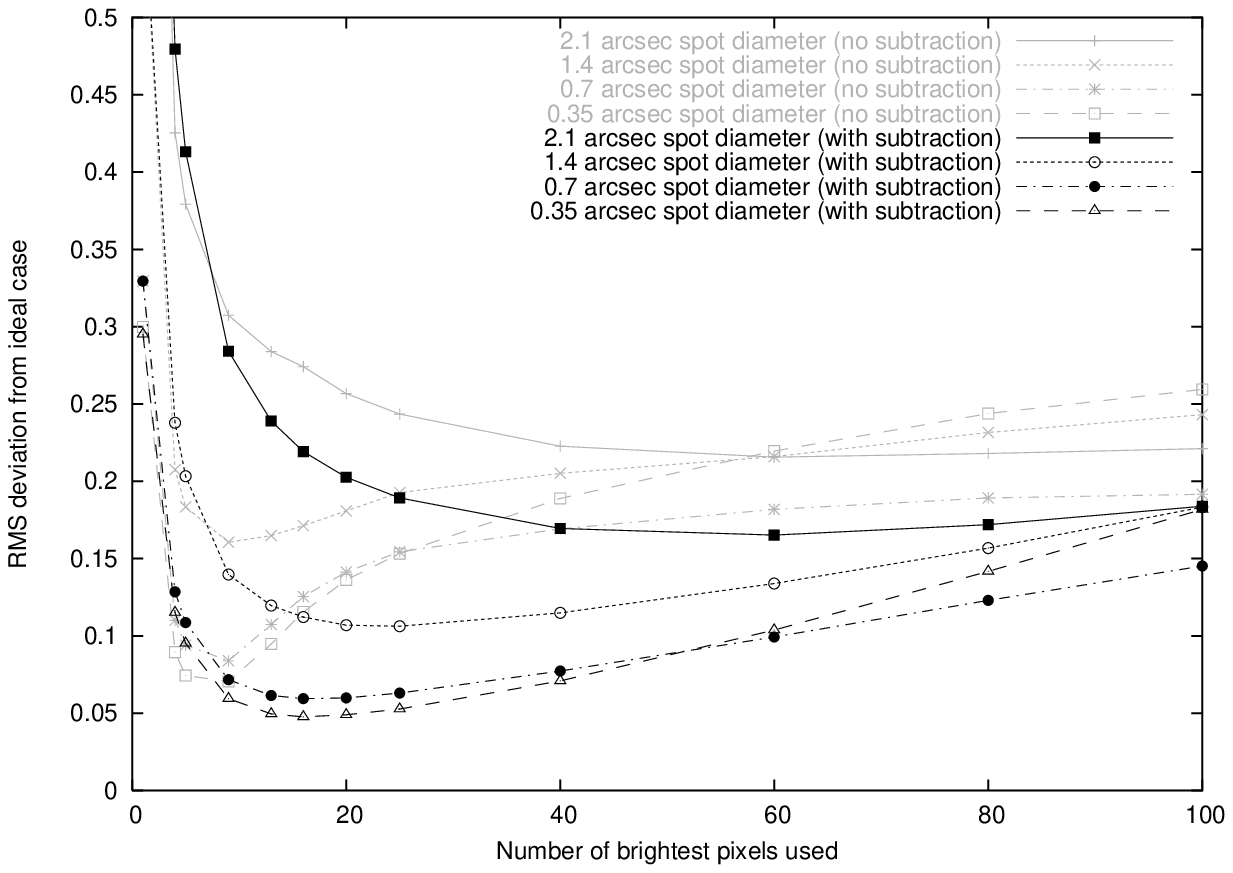}
\caption{(a) A diagram showing the Shack-Hartmann point spread
  functions used for comparison of the brightest pixel selection
  algorithm.  The spots have Airy diameters equal to 0.35, 0.7, 1.4
  and 2.1~arcsec respectively from small to large (corresponding to
  radii of 5\%, 10\%, 20\% and 30\% of the sub-aperture), normalised
  to the same maximum brightest pixel value for display purposes.  (b)
  A diagram showing an example of atmospheric broadened Shack-Hartmann point
  spread functions for comparison with (a).  (c) A plot showing slope
  calculation accuracy (lower is better) as a function of number of
  brightest pixels used for different spot radii.  Comparisons for
  pixel selection only (labelled ``no subtraction''), and using the
  ``$N+1^{th}$ subtracted algorithm'' (labelled ``with subtraction'')
  are made.  Detector readout noise and photon shot noise have been
  included.}
\label{fig:diam}
\end{figure}

The size of a Shack-Hartmann generated spot will depend on the
wavefront sensor pixel scale used and on the optical aberrations
introduced by the lenslet array and optics, as well as the seeing
conditions.  If light is concentrated into only a few pixels then
intuitively, using fewer brightest pixels is likely to give better
performance since fewer pixels will actually contain signal.
Fig.~\ref{fig:diam}(a) shows the spot \psfs used in these comparisons
and Fig.~\ref{fig:diam}(b) shows the same spots after broadening due
to atmospheric perturbation.  Fig.~\ref{fig:diam}(c) shows the slope
calculation accuracy as a function of number of brightest pixels used
for different Shack-Hartmann spot diameters (with lower results
representing greater accuracy).  It can be seen that smaller spot
sizes favour a lower number of brightest pixels used, but that this
trend is not as pronounced as might be expected from the relative
unbroadened \psf spot sizes.  However, since the atmosphere broadens
these \psfs as shown in Fig.~\ref{fig:diam}(b), they then appear more
similar in size.  It can be seen that background subtraction of the
next brightest pixel value (using the ``$N+1^{th}$ subtracted
algorithm'') is advantageous.  We find that this is in general the
case (as expected from linearity considerations), and so the rest of
the results presented in this paper show cases using the ``$N+1^{th}$
subtracted algorithm'' only, unless there is good reason to also
show results corresponding to no subtraction.  This plot suggests that
at least the 20 brightest pixels should be used along with subtraction
of the 21$^{\textrm{st}}$ brightest pixel.  By extrapolating the
curves to 256 brightest pixels, the slope estimation accuracy of the
conventional case (with no brightest pixel selection) can be seen to
be significantly poorer than using brightest pixel selection.

\subsection{Sensitivity to signal level}
\begin{figure}
\includegraphics[width=7cm]{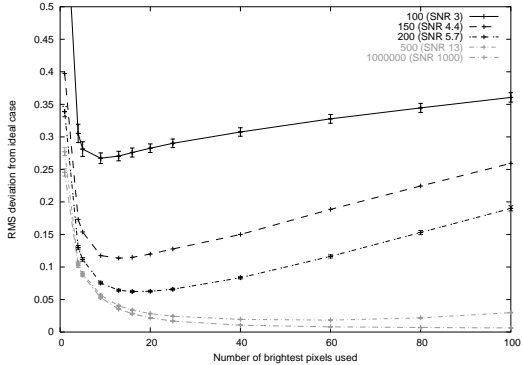}
\caption{A plot showing slope estimation accuracy as a function of
  number of brightest pixels used for different detected photon
  arrival rates, given in the key in photons per sub-aperture per
  frame.  The standard simulation parameters for this paper have been
  used, and the signal-to-noise ratio of each case (Eq.~\ref{eq:snr})
  is given in the legend.}
\label{fig:sig}
\end{figure}

The signal level received by the wavefront sensor is also a parameter
that will affect on the optimum number of brightest pixels used.
Fig.~\ref{fig:sig} shows the slope estimation accuracy (lower is
better) as a function of number of brightest pixels used for different
signal levels, with 100, 150, 200, 500 and $10^6$ (high light level)
detected photons per sub-aperture shown.  It should be noted that for
the high light level case, extrapolation to using 256 brightest pixels
(all pixels) gives better performance because the noise here is
negligible.  For other cases, using at least 20 brightest pixels (with
next-brightest subtraction) appears to give good performance at all
signal levels, though faint signals may benefit from selecting fewer
pixels.

\subsection{Sensitivity to background subtraction}
\begin{figure}
\includegraphics[width=7cm]{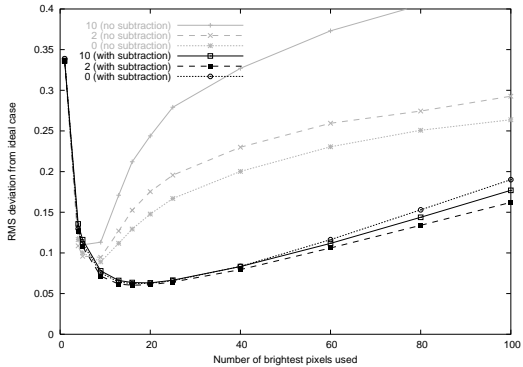}
\caption{A plot showing slope estimation accuracy as a function of
  number of brightest pixels used for different background subtraction
  errors.  The legend gives the error made in background subtraction,
  with 0 corresponding to no error, and 10 corresponding to a flat
  background of 10 counts remaining unsubtracted from the wavefront sensor
  images.  The labels ``(no subtraction)'' and ``(with subtraction)''
  refer to the brightest pixel algorithm used, either a simple
  threshold (removing all values below the $N^{th}$ brightest pixel),
  or a threshold and then subtracting the $N+1^{th}$ brightest pixel
  from unthresholded values respectively. Theoretically, the ``with
  subtraction'' curves are the same, the differences shown here being
  an artifact of the Monte-Carlo simulation.
}
\label{fig:bg}
\end{figure}

Images obtained from a \wfs are typically calibrated before
processing, which includes the removal of a background image.  This
background image must be obtained for each \wfs exposure time and
temperature to accurately remove the background.  It is often the case
that the background image used is not perfectly suited to the \wfs
images being obtained, and so systematic error results.  For
closed-loop \ao systems, this error is small since the \ao system is
trying to minimise spot deviations.  However for open-loop \ao
systems, wrong background subtraction leads to errors in wavefront
slope estimation and so the wrong command vector is sent to the
deformable mirror, reducing the \ao correction.  Here, this effect is
investigated by incorrectly removing a background and investigating
the slope estimation accuracy as the number of brightest pixels used
is changed.  We assume that $B_{\rm true} - B_{\rm subtracted} = N$
where $N$ is the same for each pixel, i.e.\ the background subtracted
image still contains a constant background equal to $N$.  Results are
shown in Fig.~\ref{fig:bg} for values of $N$ equal to one, two 
and ten counts, i.e. the background map is wrong by this many
photo-electrons.  In these cases, the true slope measurement used for
comparison (Eq.~\ref{eq:metric}) does not contain the false
background.

It can be seen that the ``$N+1^{th}$ subtracted algorithm'' gives far
better performance than the non-subtracted version because this
effectively removes the false background.  If this subtraction is not
performed then estimation accuracy worsens as the unremoved background
level is increased.  Best results are obtained here by using 20
brightest pixels with the ``$N+1^{th}$ subtracted algorithm'', far
better than is achieved without using the brightest pixel selection. 

\subsection{Sensitivity to detector readout noise}
\begin{figure}
\includegraphics[width=7cm]{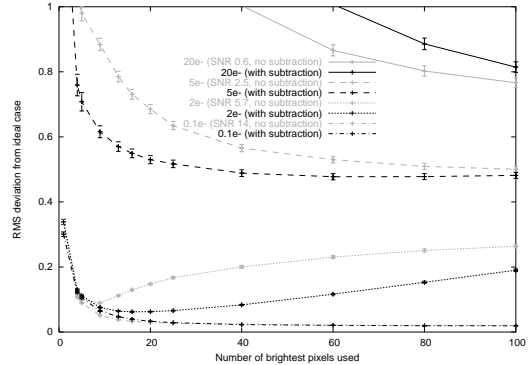}
\caption{A figure showing slope estimation accuracy as a function of
  number of brightest pixels used for cases with different CCD readout
  noise.  Comparisons for brightest pixel selection only, and using
  the ``$N+1^{th}$ subtracted algorithm'' are made.  A light
  level of 200 detected photons per sub-aperture was assumed.  The
  legend shows the signal-to-noise ratio (Eq.~\ref{eq:snr}) for each readout noise.
}
\label{fig:read}
\end{figure}

In most \ao systems, the detector readout noise of \wfss is generally
fairly low since small signal levels often require detection, and are
typically below 5 electrons.  Since readout noise is random, it
can have an impact on \ao system performance if not taken into
account.  Fig.~\ref{fig:read} shows slope estimation accuracy as a
function of number of brightest pixels used for different detector readout
noise levels.  An important point to note here is that when readout
noise is above about 5 electrons, subtracting the $N+1^{th}$ brightest
pixel value leads to poorer performance than when no such subtraction
is performed.  This is because the higher readout noise means that the
subtracted value varies significantly between frames since it contains
this noise, and therefore a significantly different background is
effectively subtracted for each frame.  It should be noted that
Fig.~\ref{fig:read} shows results for when signal level is 200 photons
per sub-aperture, while for higher signal levels the readout noise
threshold at which subtraction of the $N+1^{th}$ brightest pixel
should or should not be applied will be higher.

For cases where the readout noise is a significant contribution, it
may be possible to subtract an average of the $N+1$ to $N+M^{th}$
brightest pixels, with $M$ being a small positive integer, or by using
a temporally filtered background level for each sub-aperture
(i.e. averaging the $N+1^{th}$ brightest pixel over several frames),
which would help to remove the random effect of readout noise.
However, this is not investigated further here.  These results show
that selecting about 20 brightest pixels appears to give a reasonable
performance for most readout noise levels, though a greater number of
pixels could be advantageous when significant noise is present.  For
cases with sub-electron readout noise, such as that obtained when
using an \emccd with high gain, estimation accuracy remains almost
constant once more than about 20 pixels are selected, meaning that
brightest pixel selection is not necessary (at least in this case,
with 200 photons detected in each sub-aperture) though does no harm.

\subsection{Changes in atmospheric conditions}
\begin{figure}
(a)\includegraphics[width=7cm]{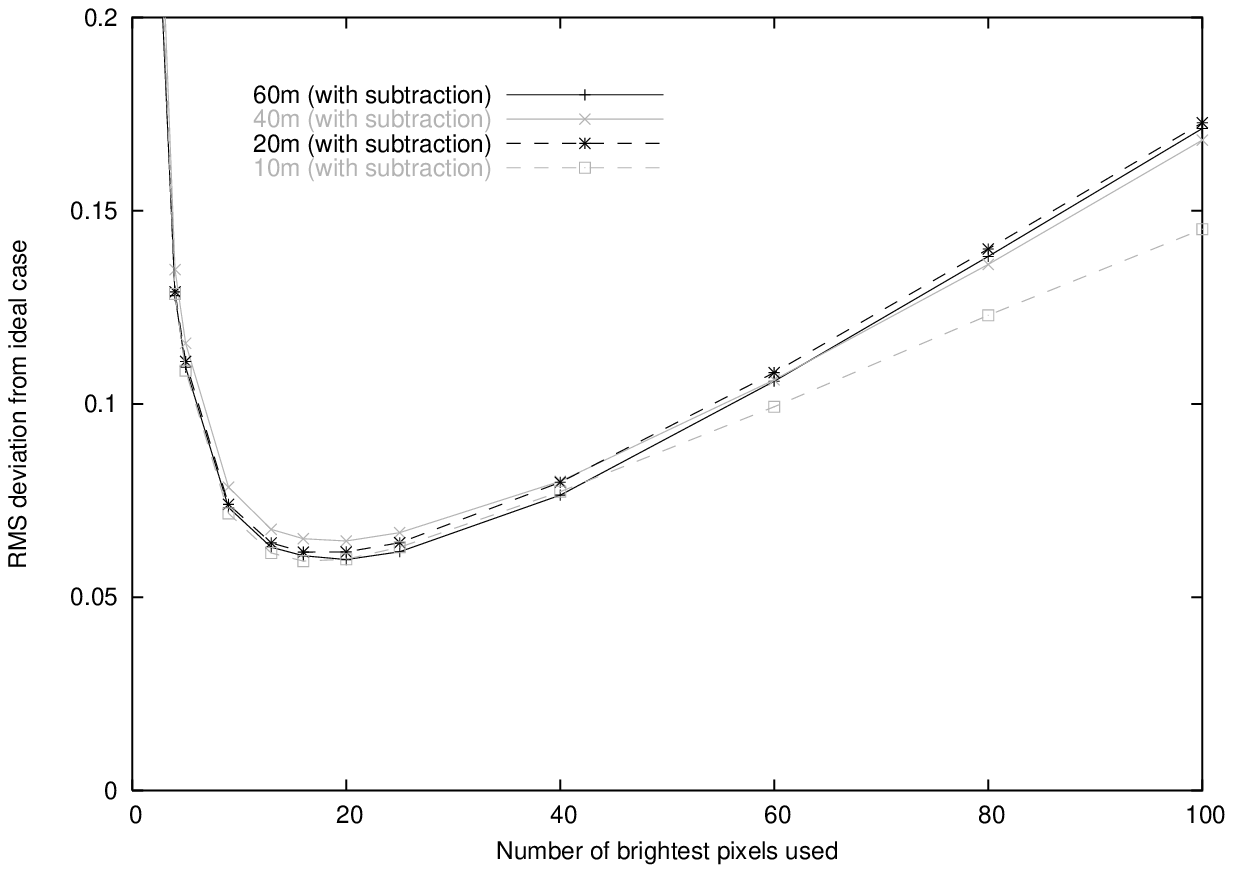}
(b)\includegraphics[width=7cm]{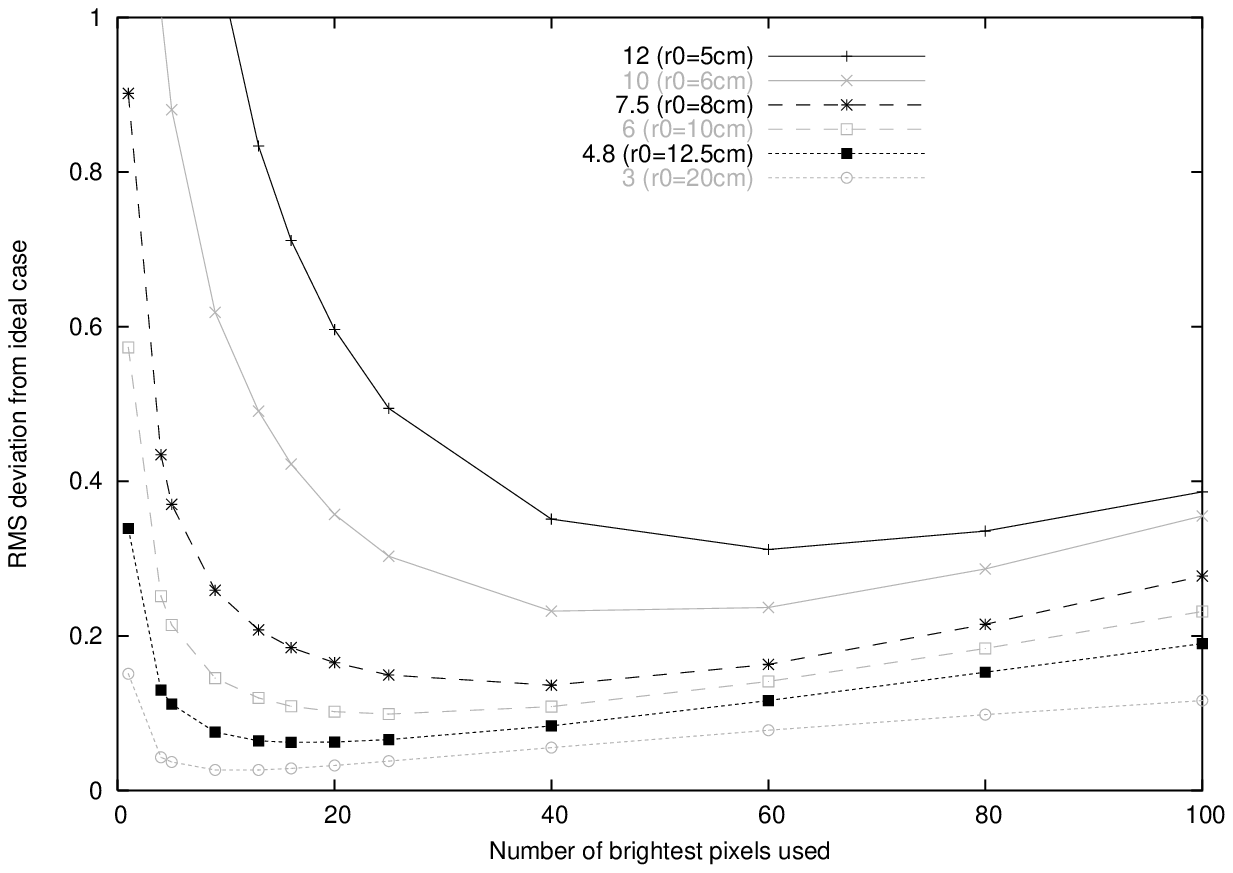}
\caption{(a) A figure showing slope estimation accuracy as a function
  of number of brightest pixels used for different atmospheric outer
  scale lengths.  (b) A figure showing
  slope estimation accuracy as a function of number of brightest
  pixels used for different atmospheric seeing conditions with $d/r_0$
  given in the legend for $d=0.6$~m.  }
\label{fig:l0}
\end{figure}

The accuracy of wavefront slope measurement is sensitive to
atmospheric conditions.  Fig.~\ref{fig:l0}(a) shows the how slope
estimation accuracy changes for different atmospheric outer scale
lengths ($L_0$).  Selection of the 20 brightest pixels
is about optimal in these cases, regardless of outer scale.

Similarly, Fig.~\ref{fig:l0}(b) shows how estimation accuracy changes
with atmospheric seeing (given by $d/r_0$ with $r_0$ being Fried's
parameter and $d=0.6$~m being the sub-aperture diameter.  Here it can
be seen that the optimal number of brightest pixels used depends
strongly on the atmospheric seeing, but that for typical values seen
during observing ($r_0=10-15$~cm), selecting 20 pixels gives good
performance (along with subtraction of the 21st pixel value).  In very
bad seeing conditions, a greater number of pixels should be selected
because the spots will be more diffuse.

\subsection{Application to laser guide star elongated spots}
So far, this paper has concentrated on application of brightest pixel
selection based on un-elongated (natural guide star) Shack-Hartmann
spots.  However, the applicability with elongated laser guide star
\citep{1985A&A...152L..29F} spots should also be considered.

\subsubsection{Rayleigh beacons}
\begin{figure}
(a)\includegraphics[width=3cm]{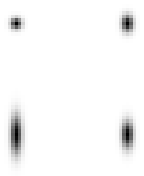}\\
(b)\includegraphics[width=7cm]{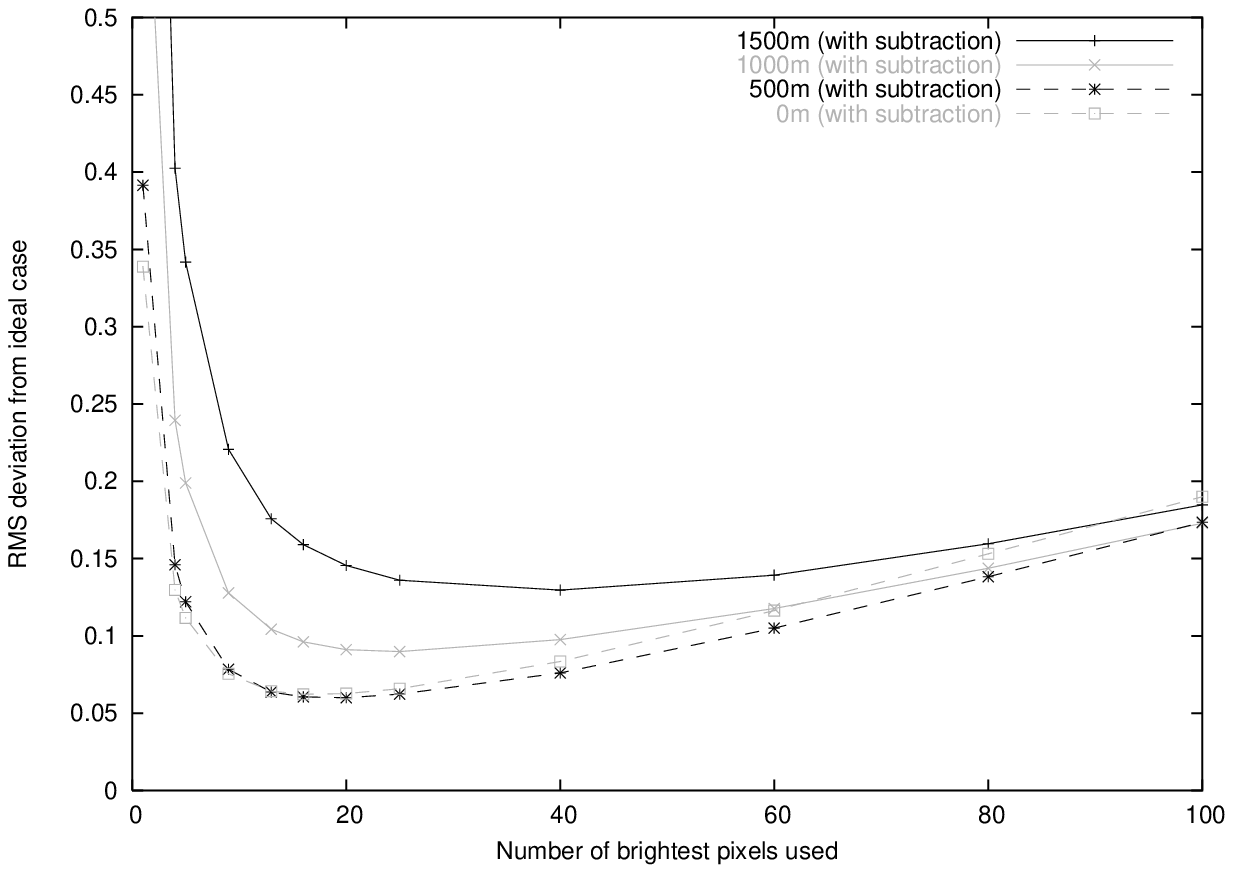}
\caption{(a) A figure showing laser guide star elongations for
  different range gate depths, these being (clockwise from top left)
  0~m, 500~m, 1000~m and 1500~m.  The guide stars are assumed to be
  Rayleigh spots at an altitude of 20~km.  (b) A figure showing slope
  estimation accuracy as a function of number of brightest pixels used
  for different laser guide star elongations.  The key shows the
  range gate depth used for each case (i.e.\ the depth of the spot).}
\label{fig:lgs}
\end{figure}

A Rayleigh laser guide star (as used with \canary) is created by
observing back scattered light from a laser beam as it transits
through the atmosphere.  By using a wavefront sensor with a fast
shutter, it is possible to observe light scattered only from a small
range of heights, thus creating a spot (artificial star) in the
atmosphere.  This spot will be geometrically elongated since it is not
viewed along the direction of propagation, and the degree of
elongation will be determined by the shutter time and the distance of
the sub-aperture from the laser launch position (when mapped into the
pupil plane).  Using elongated Shack-Hartmann spots as would be seen
by the \canary \lgs (Fig.~\ref{fig:lgs}(a)), the brightest pixel
selection algorithm is investigated.  The elongated spots shown give
the expected elongation for a centre launched Rayleigh laser beacon at
20~km with shutter times that give a light transit distance of 0, 500,
1000 and 1500~m, viewed by a 60~cm sub-aperture at the edge of a 4.2~m
telescope pupil.  The spots have an unelongated width of 0.28~arcsec.
A typical range gate depth for a Rayleigh beacon at this altitude is
between 500--1000~m, and \canary is likely to use a 600~m depth.

Fig.~\ref{fig:lgs}(b) shows the slope estimation accuracy as a
function of number of brightest pixels selected, for different degrees
of spot elongation.  This figure shows accuracy in the direction
parallel to the elongation, which has less sensitivity to wavefront
slope than perpendicular to elongation and thus represents a worst
case.  The ``$N+1^{th}$ subtracted algorithm'' gives better
performance than brightest pixel selection only and so only
subtracted results are shown.  This plot suggests that at least 20
brightest pixels should be selected, with more for very elongated
spots.  For a centre launched \lgs, spot elongation is greater for
sub-apertures at the edges of the \wfss.  Therefore, the number of
brightest pixels used may need to be adjusted according to
sub-aperture location.  Such an adjustment is possible with the \darc
\rtcs used with \canary.

\subsubsection{Sodium beacons}
\begin{figure}
(a)\includegraphics[width=7cm]{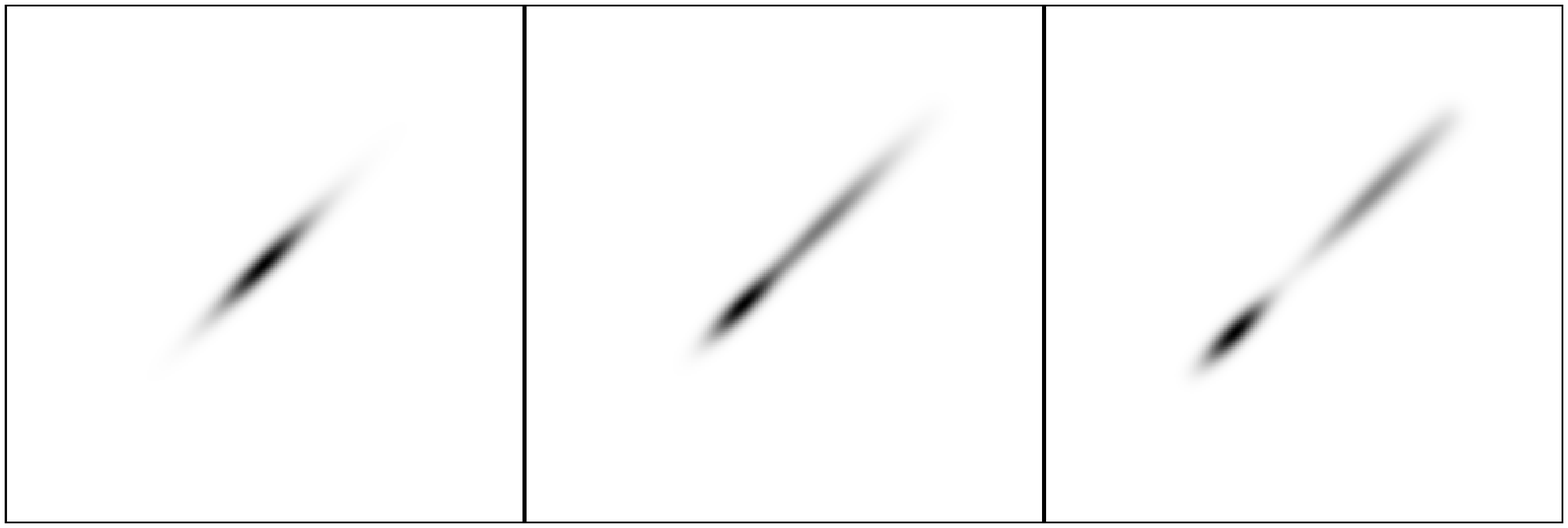}
(b)\includegraphics[width=7cm]{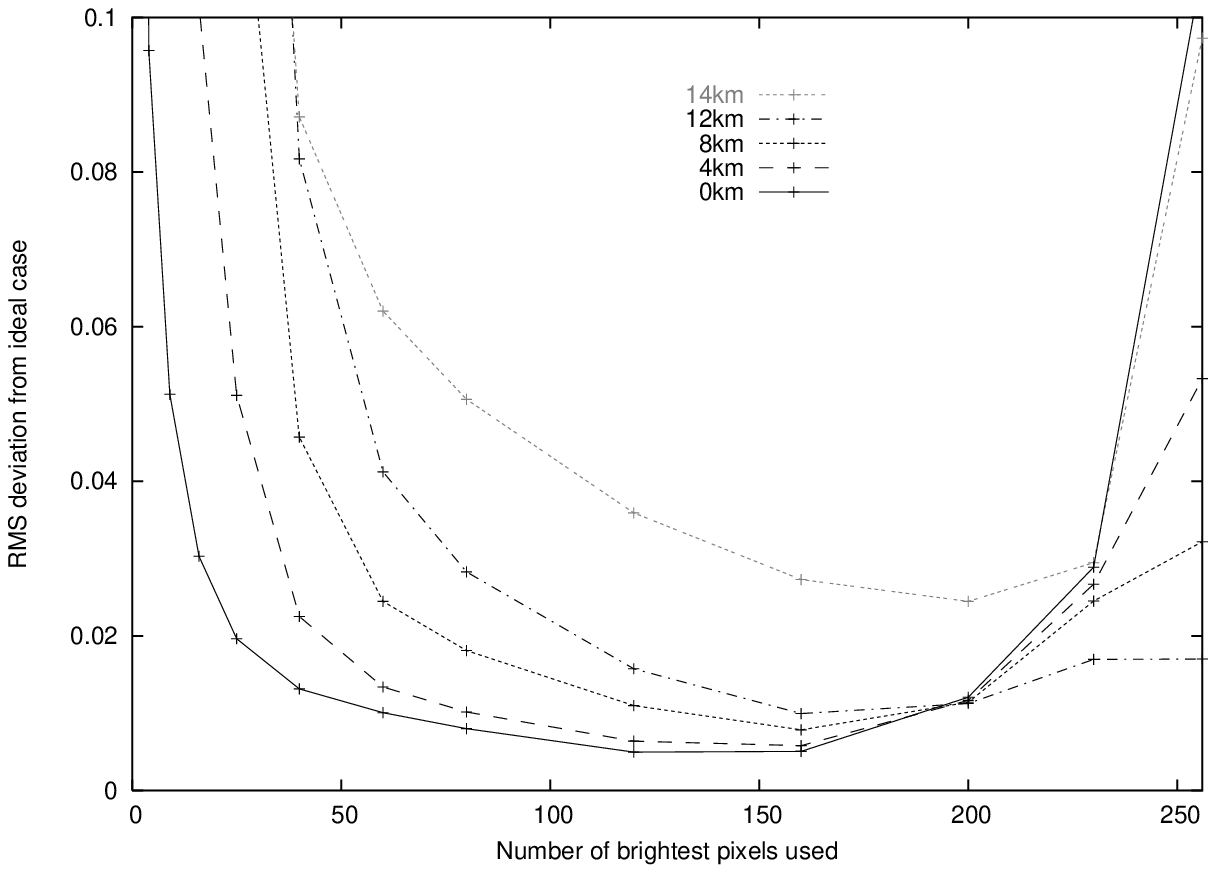}
\caption{(a) Sodium LGS sub-aperture images for different sodium profile
  peak separations (shown in negative).  Separations of 0~km, 8~km and
  14~km are shown, the double peak structure is clearly visible,
  and the surrounding box shows the sub-aperture size used here.
 (b) A figure showing slope esimation accuracy as a function of
  number of brightest pixels used for different sodium laser guide
  star models (double Gaussian model with varying peak separation).
  The legend gives the separation between the double peaks in the
  sodium profile.  Subtraction of the $N+1^{th}$ brightest pixel has
  been applied, i.e.\ the ``$N+1^{th}$ subtracted algorithm'' has been
  used.
}
\label{fig:sodium}
\end{figure}

A Sodium \lgs \citep{1985A&A...152L..29F} is created using a laser
tuned to a resonance of sodium atoms found in the mesosphere at about
90~km above the Earth's surface.  These atoms absorb laser light and
then re-emit light in all directions when they decay, thus creating a
signal that is detectable at the telescope.  The density profile (with
respect to altitude) of these atoms is known to vary with time
\citep{SodiumProfileData} meaning that the structure of the returned
signal varies.  As with a Rayleigh \lgs, the spots will be
geometrically elongated with the degree of elongation increasing with
the distance between the laser launch telescope and the wavefront
sensor sub-aperture in question.  For the purposes of investigation
here, we model the sodium layer as having a double Gaussian profile
\citep{SodiumDoublePeak}, with one peak at 90~km and the other at a
separation ($s$) above this, with $s$ ranging from 0--14~km.  We
assume peak sodium densities of 40\% and 60\% respectively with peak
half-widths of 3.5 and 2~km.  We consider the case of a sub-aperture
placed 21~m from the laser launch telescope position with a 24~arcsec
field of view, which gives elongations as shown in
Fig.~\ref{fig:sodium}(a), with a $45^\circ$ rotation for display
purposes, and assume a light level of 1000 detected photons per
sub-aperture per frame.  To investigate the performance of the
brightest pixel selection algorithm, we show here results for which we
are measuring wavefront slope in a direction parallel to spot
elongation, which represents the worse case (least sensitive)
scenario.

Fig.~\ref{fig:sodium}(b) shows the slope estimation accuracy as a
function of number of brightest pixels selected using the ``$N+1^{th}$
subtracted algorithm'', obtained using the
method presented in \S~\ref{sect:thealgo} for different double peaked
sodium profiles.  The presence of structure in the sodium spots means
that these spots require greater numbers of brightest pixels to be
selected than has previously been used to achieve best performance.
Fig.~\ref{fig:sodium}(b) shows that selecting more than 100 brightest
pixels in this case gives best performance, with more required when
elongation increases.  Elongation can be due to both the sodium layer
profile and (as discussed previously) because of increasing distance
of the sub-aperture from the laser launch position (increasing
geometrical elongation).  These results suggest that when using sodium
laser guide stars, brightest pixel selection is beneficial (compared
with the case for 256 pixels which corresponds to no selection),
though care must be taken to select enough pixels.

\subsection{Application to different sub-aperture sizes}
\begin{figure}
\includegraphics[width=7cm]{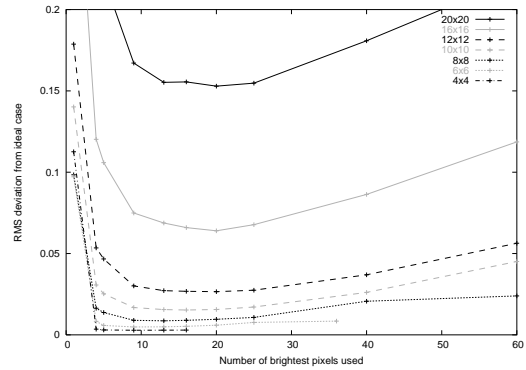}
\caption{A figure showing slope estimation accuracy as a function of
  number of brightest pixels selected for different sub-aperture sizes
  (number of pixels per sub-aperture) as shown in the key.
  Subtraction of the next brightest pixel has been made.}
\label{fig:nimg}
\end{figure}

Open-loop \ao systems will generally have more pixels per sub-aperture
than closed-loop systems, because spot motions are larger since the
\wfs does not see corrections made by the \dm.  So far, we have
considered the case with $16\times16$ pixels per sub-aperture, however
it is interesting to see how brightest pixel selection performs with
different numbers of pixels.  Fig.~\ref{fig:nimg} shows performance
for different sub-aperture sizes as a function of number of brightest
pixels used.  This shows that for smaller sub-apertures, selecting
fewer brightest pixels gives better performance, though if too few are
selected, performance will be significantly degraded.  For the
smallest sub-apertures, there is little advantage in implementing the
brightest pixel selection algorithm.

It should be noted that although the curves representing sub-apertures
with fewer pixels are lower, this does not mean that this will give
better performance; rather it shows the performance that can be achieved
relative to the noiseless case.

\subsection{Linearity}
\begin{figure}
(a)\includegraphics[width=7cm]{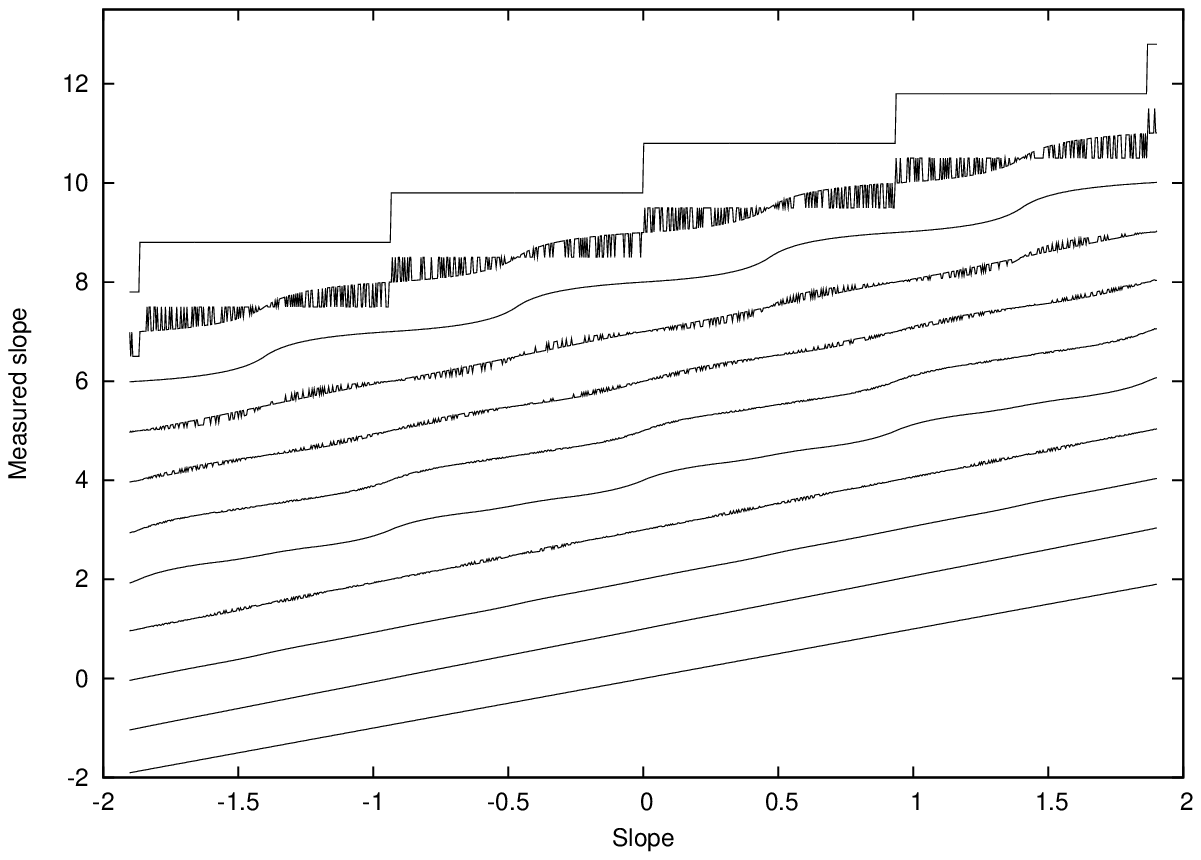}
(b)\includegraphics[width=7cm]{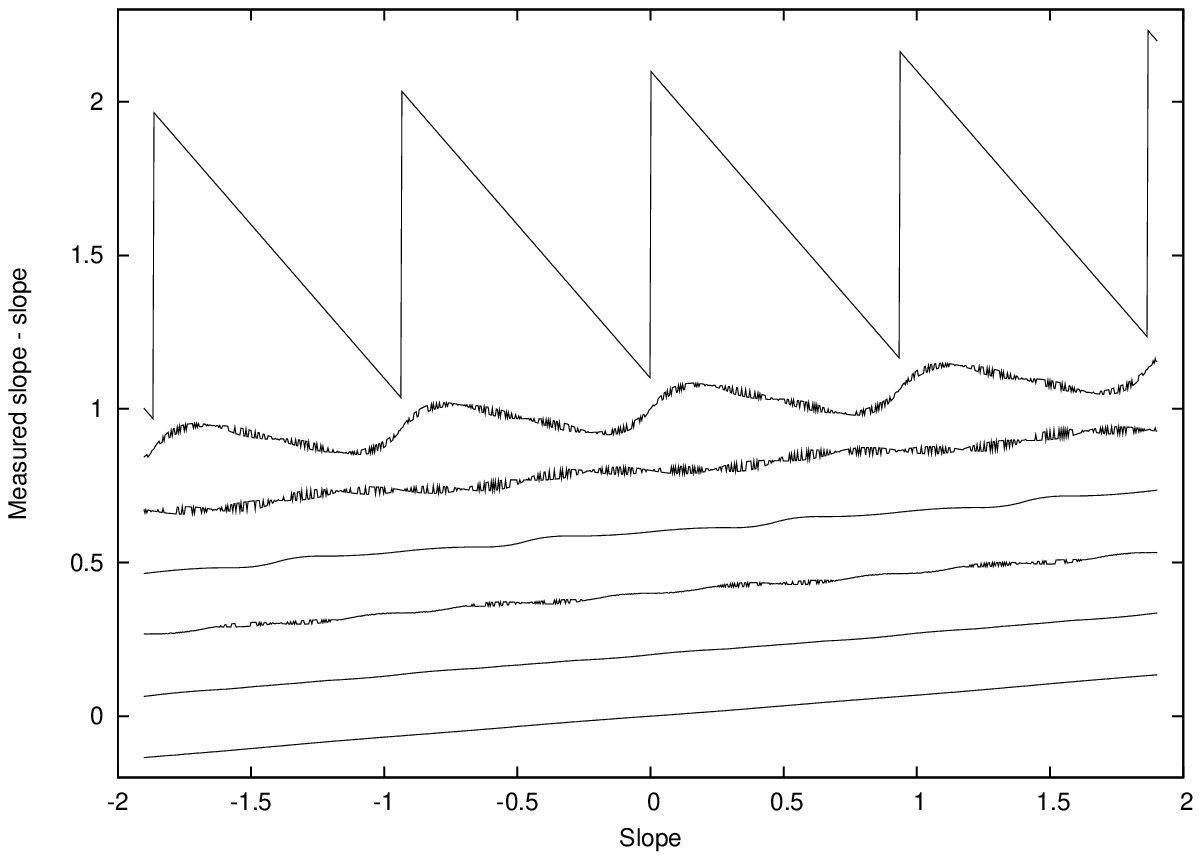}
(c)\includegraphics[width=7cm]{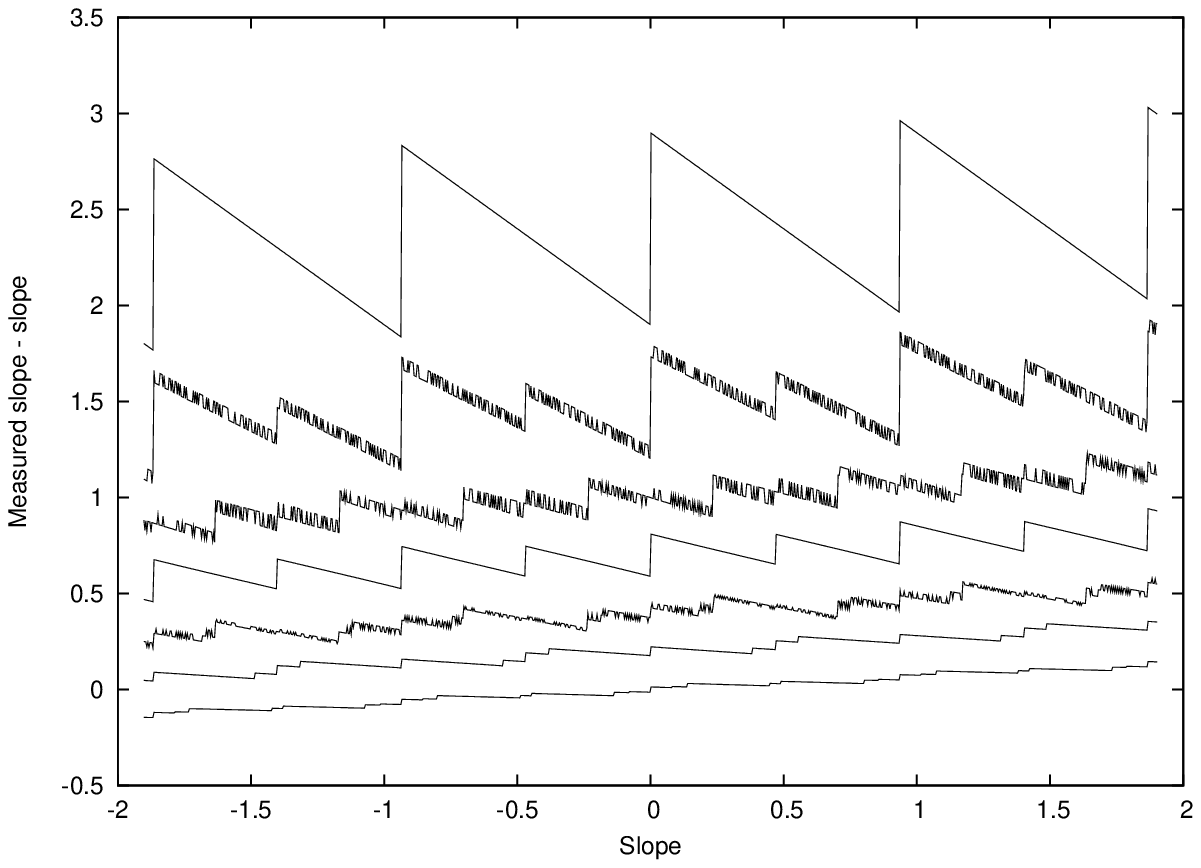}
\caption{(a) A figure showing wavefront slope measured using brightest
  pixel selection and subtraction of the next brightest pixel against
  the slope measured using all pixels (no selection).  When few pixels
  are used, there are notable steps in slope estimation.  Each trace
  has been offset vertically for clarity.  From top to
  bottom, the number of brightest pixels are 1, 3, 4, 5, 7, 9, 10, 13,
  20, 50, all (256).  (b) A figure showing wavefront slope measured
  using brightest pixel selection and subtraction of the next
  brightest pixel after subtraction of slope measured
  using all pixels, against the slope measured using all pixels.  Each
  trace has been offset vertically for clarity.  From top
  to bottom the number of brightest pixels selected are 1, 9, 15, 20,
  25, 50 and 100.  (c) As for (b) but using brightest pixel selection
  only, without subtraction of the next brightest pixel.  }
\label{fig:linearity}
\end{figure}

Fig.~\ref{fig:linearity}(a) shows estimated wavefront slope across a
noiseless sub-aperture using the brightest pixel selection algorithm
for different numbers of selected pixels, as the slope across the
sub-aperture is changed.  Here, the ``$N+1^{th}$ subtracted algorithm''
has been implemented.  It can be seen that there is significant
non-linearity when small numbers of brightest pixels are selected,
exemplified in the case using only the single brightest pixel.  In
this case, as the wavefront is gradually tilted causing the spot to
move gradually across the sub-aperture, the brightest pixel will jump
at fixed intervals from one pixel to the next, causing a jump in the
corresponding slope estimation.  When selecting more than one pixel,
these jumps still occur, but as more and more pixels are selected the
jumps become less profound.  By the time 20 brightest pixels are
selected, the change in estimated slope is almost linear.

Fig.~\ref{fig:linearity}(b) shows the estimated wavefront slope after
subtraction of the slope calculated using all pixels, to show
non-linearities more clearly.  It should be noted that even with
selection of 100 brightest pixels, the trace is not horizontal.  This
is due to the finite nature of the sub-aperture and the infinite size
of the spot when all pixels are selected (an Airy disc pattern does
not have a cut off beyond which all pixels are zero).  Fortunately,
because these traces are linear this is an effect that can be taken
into account during the wavefront reconstruction (for example by
scaling a control matrix by a small factor, or simply creating the
control matrix using the same number of selected brightest pixels).
Therefore, provided a large enough number of brightest pixels are
selected (for example at least 20--25 in this case), the performance of
the \ao system will not be degraded.

Fig.~\ref{fig:linearity}(c) shows the linearity of the brightest pixel
selection algorithm without subtraction of the $N+1^{th}$ brightest
pixel, and this is considerably more non-linear than
Fig.~\ref{fig:linearity}(b), suggesting that this subtraction should
always be applied.

\subsection{Computational complexity}
\label{sect:computational}
The brightest pixel selection algorithm implemented in \darc for
\canary uses a quick-sort sorting algorithm, which on average makes
$O\left( n \log n\right)$ comparisons (with a worst case $n^2$
comparisons), with $n$ being the number of pixels in a sub-aperture.
This is repeated each iteration for each sub-aperture and so increases
the \ao system latency (the time between \wfs readout and the \dm
shape being modified).

We have measured the latency when using and not using brightest pixel
selection for the standard \canary phase A configuration.  We find
that using brightest pixel selection adds about 330~$\mu$s to the
latency of the system.  These measurements were made using the
software timing functions within \darc.  Due to the configurable
nature of \darc, we were able to optimise the number of processing
threads used in each case, and optimum performance is obtained by
increasing from 14 to 16 the number of threads when switching from no
brightest pixel selection to brightest pixel selection.

When using brightest pixel selection, altering the number of pixels
selected, and including subtraction of the next brightest does not
affect the latency significantly, since most of the computation is in
sorting the pixels which is performed regardless of the number of
pixels used.

For \canary, this latency increase does not significantly affect
performance because latency is found to contribute only a small part
of the overall error budget.  However, for other systems, particularly
higher order systems, this may have a significant affect, and so the
benefits of brightest pixel selection should be weighted against the
effect of increased latency.  

\subsection{Wavefront error reduction}
The main purpose of an \ao system is to remove residual wavefront
error.  For the default parameter set studied here, the results
presented show an improvement of about 0.2 pixels \rms in slope
estimation accuracy when comparing the pixel selection and traditional
(no brightest pixel selection) algorithms.  With a pixel scale of
0.22~arcsec per pixel this corresponds to 0.044~arcsec improvement,
or, by considering phase across a 0.6~m diameter sub-aperture, about 130~nm of
wavefront error.

We have performed full end-to-end Monte-Carlo simulations of CANARY
using the \dasp \citep{basden5} and including the brightest pixel
selection algorithm.  These simulations included three off-axis
natural guide star wavefront sensors with locations specified in
table~\ref{table:asterism}, and a \dm in open-loop correcting an
on-axis science path.  With a 250~Hz update rate, we assume 200
detected photons per sub-aperture per frame (the default for this
paper), and also assume a detector with two electron read noise,
corresponding to an \emccd with moderate multiplication gain.  These
parameters are chosen to match the default in the rest of this paper
and to clearly show the advantage given by brightest pixel selection.
The atmosphere model has a global $r_0$ of 12.5~cm and matches data
taken on-sky with \canary, namely, a strong ground layer (60\% at 0m
and 30\% at 500~m), and a weaker layer (10\% strength) at 5~km.
Least-squares tomographic wavefront reconstruction was used by
reconstructing phase at the layer heights, and then projecting phase
down onto the on-axis open-loop \dm.

\begin{figure}
\begin{tabular}{lll}
\hline
Wavefront sensor & Radial offset & Angular offset \\
& / arcsec & / degrees\\ \hline
1 & 40.2 & 26.6\\
2 & 40.2 & 153.4\\
3 & 19.0 & 288.4\\
Science & 0 & 0\\ \hline
\end{tabular}
\caption{Position of guide stars relative to on-axis location in a
  polar coordinate system.}
\label{table:asterism}
\end{figure}

\begin{figure}
\includegraphics[width=8cm]{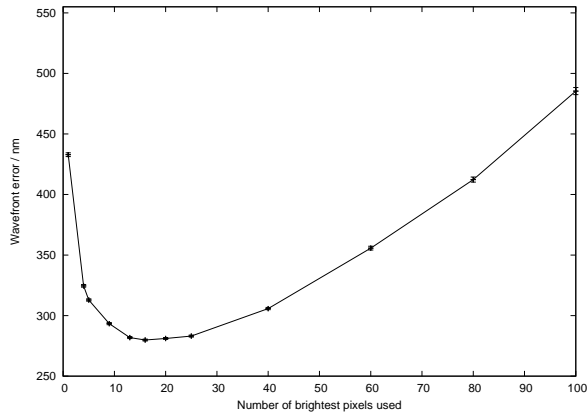}
\caption{A figure showing residual wavefront error for an open-loop
  MOAO simulation based on CANARY, as a function of number of
  brightest pixels selected.  Without brightest pixel selection, the
  error is about 550~nm.}
\label{fig:daspresults}
\end{figure}

Fig.~\ref{fig:daspresults} shows the estimated \canary residual
wavefront error as a function of number of brightest pixels selected
(using the ``$N+1^{th}$ subtracted algorithm'').  It can be seen
that brightest pixel selection can greatly reduce wavefront error,
particularly in cases like this with a low or moderate signal-to-noise
ratio.  Without brightest pixel selection the simulation residual
wavefront error is about 550~nm which gives a Strehl ratio of 6\%,
dominated by noise and tomographic error (a full \canary error budget
is given by \citep{canaryresultsshort}).  When 20 brightest pixels are
selected the Strehl ratio increases to 33\%, and the \rms \wfs is
reduced by about 270~nm to 280~nm.  Therefore, for open-loop \ao
systems with large numbers of pixels per sub-aperture, brightest pixel
selection can lead to significant performance improvement.

\section{Conclusion}
The application of a threshold based on the $N^{th}$ brightest pixel
in each sub-aperture of a Shack-Hartmann wavefront sensor has been
investigated, with performance being studied for a wide range of
parameters, with particular emphasis on linearity for application to
open-loop \ao, for example \moao.  This algorithm involves selecting
the $N$ brightest pixels in each sub-aperture and setting to zero all
pixels below this level.  Additionally, it has been shown here that
there is a significant gain in slope estimation accuracy and
linearity if the $N+1^{th}$ pixel value is subtracted from all
non-zero pixels.

From the results presented here, using the 20 brightest pixels (with
subtraction of the 21$^{\textrm{st}}$ brightest pixel seems optimal
for average seeing, while more pixels, up to 40, should be used for
poorer seeing, dependent on signal level and detector noise.  One
should always use more than about 10 brightest pixels otherwise
performance will be significantly degraded.  This technique is suited
to Shack-Hartmann wavefront sensors with at least $4\times4$ pixels
per sub-aperture, and significant reductions in wavefront error can be
achieved.  Elongated laser guide star spots also require selection of
a larger number of pixels.

\section*{Acknowledgements}
This work is funded by the UK Science and Technologies Facility Council. 
\bibliographystyle{mn2e}

\bibliography{mybib}
\bsp

\end{document}